\documentclass{aa}  

\usepackage{graphicx}
\usepackage{txfonts}
\begin{document} 
\title{The extinction and dust-to-gas structure of the planetary nebula 
NGC~7009 observed with MUSE\footnote{Based on observations collected at the 
European Organisation for Astronomical Research in the Southern Hemisphere 
under ESO programme 060.A-9347(A).}
}
\titlerunning{Extinction and dust in NGC~7009}

\author{J. R. Walsh\inst{1}
        \and
        A. Monreal-Ibero\inst{2}   
        \and
        M. J. Barlow\inst{3}
        \and
        T. Ueta\inst{4} 
        \and
        R. Wesson\inst{5}
        \and
        A. A. Zijlstra\inst{6}
}

 \institute{European Southern Observatory. Karl-Schwarzschild Strasse 2,  D-85748 Garching, 
 Germany \\
            \email{jwalsh@eso.org}
            \and
            GEPI, Observatoire de Paris, PSL Research University, CNRS, Universit\'e Paris-Diderot, Sorbonne Paris Cit\'e, Place Jules Janssen, 92195 Meudon, France 
%            \email{Ana.Monreal-Ibero@obspm.fr}
            \and
            Dept. of Physics and Astronomy, University College London, Gower Street, London WC1E 6BT, United Kingdom 
%            \email{mjb@star.ucl.ac.uk}
            \and
            Department of Physics and Astronomy, University of Denver, 2112 E. Wesley Ave., Denver, CO, 80210, USA 
%            \email{Toshiya.Ueta@du.edu}                          
            \and
            European Southern Observatory, Alonso de C\'{o}rdova 3107, Casilla 19001, Santiago, Chile 
%            \email{rwesson@eso.org} 
            \and
            Jodrell Bank Centre for Astrophysics, Alan Turing Building, University of Manchester, Manchester M13 9PL, UK 
%            \email{albert.zijlstra@manchester.ac.uk} 
} 

\date{Received: 17 December 2015; accepted: 17 February 2016}
\authorrunning{Walsh, J. R., et al.}

\abstract
% context heading (optional)
{Dust plays a significant role in planetary nebulae. Dust ejected with the gas 
in the asymptotic giant branch (AGB) phase is subject to the harsh environment of the 
planetary nebula (PN) while the star is evolving towards a white dwarf. Dust surviving the 
PN phase contributes to the dust content of the interstellar medium.}
% aims heading (mandatory)
{The morphology of the internal dust extinction has been mapped for the first time 
in a PN, the bright nearby Galactic nebula NGC~7009. The morphologies of the gas, 
dust extinction and dust-to-gas ratio are compared to the structural features of the 
nebula.}
% methods heading (mandatory)
{Emission line maps in H Balmer and Paschen lines were formed from analysis of
MUSE cubes of NGC~7009 observed during science verification of the instrument. The 
measured electron temperature and density from the same cube
were employed to predict the theoretical H line ratios and derive the extinction 
distribution across the nebula. After correction for the interstellar extinction 
to NGC~7009, the internal $A_V/N_{H}$ has been mapped for the first time in a PN.}
% results heading (mandatory)
{The extinction map of NGC~7009 has considerable structure, broadly corresponding 
to the morphological features of the nebula. The dust-to-gas ratio, $A_V/N_{H}$, 
increases from 0.7 times 
the interstellar value to $>$5 times from the centre towards the periphery of the 
ionized nebula. The integrated $A_V/N_{H}$ is about 2$\times$ the mean ISM value. A 
large-scale feature in the extinction map is a wave, consisting of a crest and trough, 
at the rim of the inner shell. The nature of this feature is investigated and 
instrumental and physical causes considered; no convincing mechanisms were
identified to produce this feature, other than AGB mass loss variations.
}
% conclusions heading (optional), leave it empty if necessary 
{Extinction mapping from H emission line imaging of PNe with MUSE provides
a powerful tool for revealing the properties of internal dust and
the dust-to-gas ratio. 
% The mature PN NGC~7009 reveals a double shell 
% dust structure as well as smaller scale dust features.
}

\maketitle
%
%________________________________________________________________

\section{Introduction}

\object{NGC~7009} (PNG 037.7 $-$34.5) is called the Saturn Nebula
since it resembles Saturn and its rings. It can be 
considered as a classic high ionization planetary nebula 
(PN), and its very high surface brightness and spatial extent, about 50$''$, 
have ensured it is well observed from the X-ray to the radio. Its optical
appearance is differentiated into an inner bright core with a sharply delineated 
elliptical rim, an outer elliptical shell, a more circular halo whose brightness 
falls off rapidly, and two extensions (streams) from the ends of the major axis 
ending in stubs called ansae. \citet{Sabbadin2004} name these morphological structures 
as main (inner) shell, outer shell, halo, streams, and ansae without attributing any
kinematical implications to this terminology. These structures are also differentiated
in emission depending on the ionization and critical density of the emission species, as 
shown by emission line mapping studies (\citet{Bohigas1994}; \citet{Phillips2010}). Several
other sub-structures are distinguished such as the two caps extending across the ends of
the major axis and the equatorial pseudo-ring along the minor axis (terminology of 
\citet{Sabbadin2004}). \citet{Balick1998} suggest that the caps and
ansae are FLIERs (fast, low-ionization emission regions) but there is no evidence for
their abundances differing from the bulk of the nebula \citep{Goncalves2003}.

The central star of spectral type O with H\,I and He\,II absorption lines has
an effective temperature of 82\ 000K (\citet{Mendez1988}; \citet{Mendez1992}) and a 
luminosity of log $L \simeq 3.70L_{\odot}$ for a 
distance of 1400 pc (adopted from \citet{Sabbadin2004}). Thus the star
is on the horizontal part of the post-AGB track, implying a mature nebula of a few 
thousand years. The velocity field has been modelled (\citet{Sabbadin2004}
and \citet{Steffen2009}), and the expansion velocity of the inner shell is
$\sim$ 22 kms$^{-1}$, but up to 35 kms$^{-1}$ in the outer shell \citep{Sabbadin2004},
although these models express the expansion as a function of projected distance from the
central star. The ansae expand faster, with VLA proper motions of 30 mas yr$^{-1}$ averaged
for both ansae \citep{Rodriguez2007}, equivalent to 200 kms$^{-1}$ at 1.4 kpc. This qualifies 
them as genuinely fast with respect to the expansion of the inner and outer shells.

The extinction to the nebula has been measured in many works and is generally low, with quoted
values in the range $0.08 < c < 0.20$; the interstellar extinction in the direction of NGC~7009 
is $c$ = 0.13 based on the reddening maps of \citet{Schlafly2011} for $E_{B-V}$ = 0.090 mag. 
Dust continuum has been measured in the infrared with ISO and a strong
excess between 28 and 44 $\mu$m has been attributed to crystalline silicate 
\citep{Phillips2010}, in agreement with the C/O ratio of $\sim$ 0.5 (\citet{Liu1995};
\citet{Fang2013}). Diffuse X-ray emission indicative of a gas at $\sim 2 \times 10^6$K 
is detected in the central core, within the inner shell, by XMM-Newton 
\citep{Guerrero2002} and Chandra \citep{Kastner2012}, and an X-ray point source 
of median energy 0.7 keV is located at the position of the central star observed by Chandra
\citep{Kastner2012}. \citet{Tsamis2008} performed integral field spectroscopy over a section
covering the inner shell and derived maps of physical diagnostics ($N_{\rm e}$, $T_{\rm e}$) and
chemical abundances including recombination line abundances. NGC~7009 has a mild abundance
discrepancy factor (ADF) between the collisional and recombination line abundances of 
3-5, depending on the ion (\citet{Liu1995}; \citet{Fang2013}).

NGC~7009 was observed in the science verification programme of the Multi Unit Spectroscopic 
Explorer (MUSE) mounted on VLT UT4 as a good test of the instrument capabilities to accurately 
map emission lines with a flux ratio over a range of 10$^{5}$. Full details of the observations,
data reduction and presentation of emission line maps, diagnostics and abundances will be 
presented in a following paper (hereafter Paper II). The topic of the current paper is the 
extinction across the nebula revealed by these MUSE observations and a study of its diverse 
structures.

%__________________________________________________________________

\section{MUSE observations and reductions}
NGC~7009 was observed with MUSE on the nights of 22 and 24 June 2014 as part of the science 
verification. The standard wide field mode was used, with a field of 60$\times$60$''$, 
0.2$''$ spaxel size and wavelength coverage  4750 - 9300\AA\ at a mean spectral resolution of 
$\sim$2500. For each spatial dither, exposures of 10, 60 and 120s were specified to ensure a 
range of the brightest emission lines were unsaturated in each exposure. The field centre was set 
to the coordinates of the central star (21$^{h}$ 04$^{m}$ 10$^{s}$.76,  
$-$11$^\circ$ 21$'$ 48$''$.6, J2000) but a position angle of 33$^{\circ}$ was specified 
so that the long axis of the nebula, including the ansae, would lie along the diagonal 
of the field. For each exposure a set of 5 dithers consisting of a central pointing and 
four 0.6$''$ shifts were employed and a second identical sequence of exposures with a 
90$^{\circ}$ rotation of the instrument position angle. A sky position at 150$''$ 
offset in PA 50$^{\circ}$ was also observed for a sky subtraction frame. Full details 
of the observing set-up will be provided in Paper II (in prep.). 

The MUSE datasets were reduced with the instrument pipeline version 1.0 \citep{Weilbacher2014}
using the delivered calibration frames (bias, flat field, arc lamp exposures) and the
pipeline astrometric, atmospheric extinction and flux calibration files. 
A single combined cube for each exposure level was produced, taking care to align the 
spatial pointing for all the data sets (using the prominent central star of NGC~7009). 
The standard pipeline sampling of 0.2$''$ and 1.25\AA\ was employed for the combined cubes.
As measured from the central star, the FWHM of the
final cubes did not exceed 0.60$''$ even at the shortest wavelength measured (a synthesized 
band free of strong emission lines centred on 4900\AA). 

Emission line maps were produced by automatically fitting Gaussians to the prominent
emission lines of H, He, N, O, S and Cl using a custom code which fits the continuum,
excluding the emission lines, by a cubic spline. Maps of the flux of the various emission 
lines were formed. The statistical errors produced by the MUSE pipeline were propagated 
to the fitted line fluxes and corresponding flux error maps were also produced.

%__________________________________________________________________

\section{Derivation of extinction map}
The classical method to measure the extinction to a planetary nebula is from the observed
ratios of the H Balmer lines, given that the relative strengths of these lines vary
weakly with temperature. Using the tabulating code of \citet{Storey1995}, the Case B 
H$\alpha$/H$\beta$ ratio varies by about 10\% over the temperature range 5000 to 20000K and
changes by 2\% over the density range 100 -- 10$^{6}$ cm$^{-3}$.
Previous measurements of the electron temperature $T_{\rm e}$ in NGC~7009, such as from the
[O~III] 5007/4363\AA\ ratio, indicate values close to 10\,000K (e.g., \citet{Aller1975}; 
\citet{Aller1977}; \citet{Czyzak1979}; \citet{Barker1983}; \citet{Hyung1995A}; 
\citet{Hyung1995B}; \citet{Rubin2002}; \citet{Goncalves2003}; \citet{Sabbadin2004}; 
\citet{Phillips2010}; \citet{Fang2011}) with little
variation. \citet{Rubin2002} determined a mean temperature averaged over the inner and
outer shells and ansae of 9912K and a fractional mean square $T_{\rm e}$ variation of 
$\lesssim$ 0.01. \citet{Phillips2010} mapped the electron density from the [S II]6716/6731\AA\
ratio and the mean is about 4000 cm$^{-3}$ in the inner and outer shells, in agreement with
determinations from other $N_{\rm e}$ sensitive line ratios for small aperture/slit spectra 
(e.g., \citet{Fang2011}). 

A first extinction map was produced from the 10s H$\alpha$ and 120s H$\beta$ maps.
The ratio of the 10s and 120s H$\beta$ flux images showed absolute agreement at the
1\% level so these can be considered interchangeable, but the longer exposure image was 
preferred for its higher signal-to-noise. The 60s and 120s H$\alpha$ images are both 
saturated in many spaxels - the line peak is saturated (top hat profile) and Gaussian fits 
are a poor representation of the true flux. Constant values of 10$^{4}$K and 4000cm$^{-3}$ 
were assumed to form a map of the extinction, c, using the Galactic law (\citet{Seaton1979};
\citet{Howarth1983}) with R = 3.1. The logarithmic extinction at H$\beta$ is 
related to reddening E$_{B-V}$ by $c$ = 1.45\,E$_{B-V}$ and to $A_{V}$ by $c$ = 0.47\,$A_{V}$ for 
this reddening law (see \citet{Howarth1983}), so the term "extinction" is used synonymously. 

\begin{figure*}
\centering
\resizebox{\hsize}{!}{
\includegraphics[angle=0]{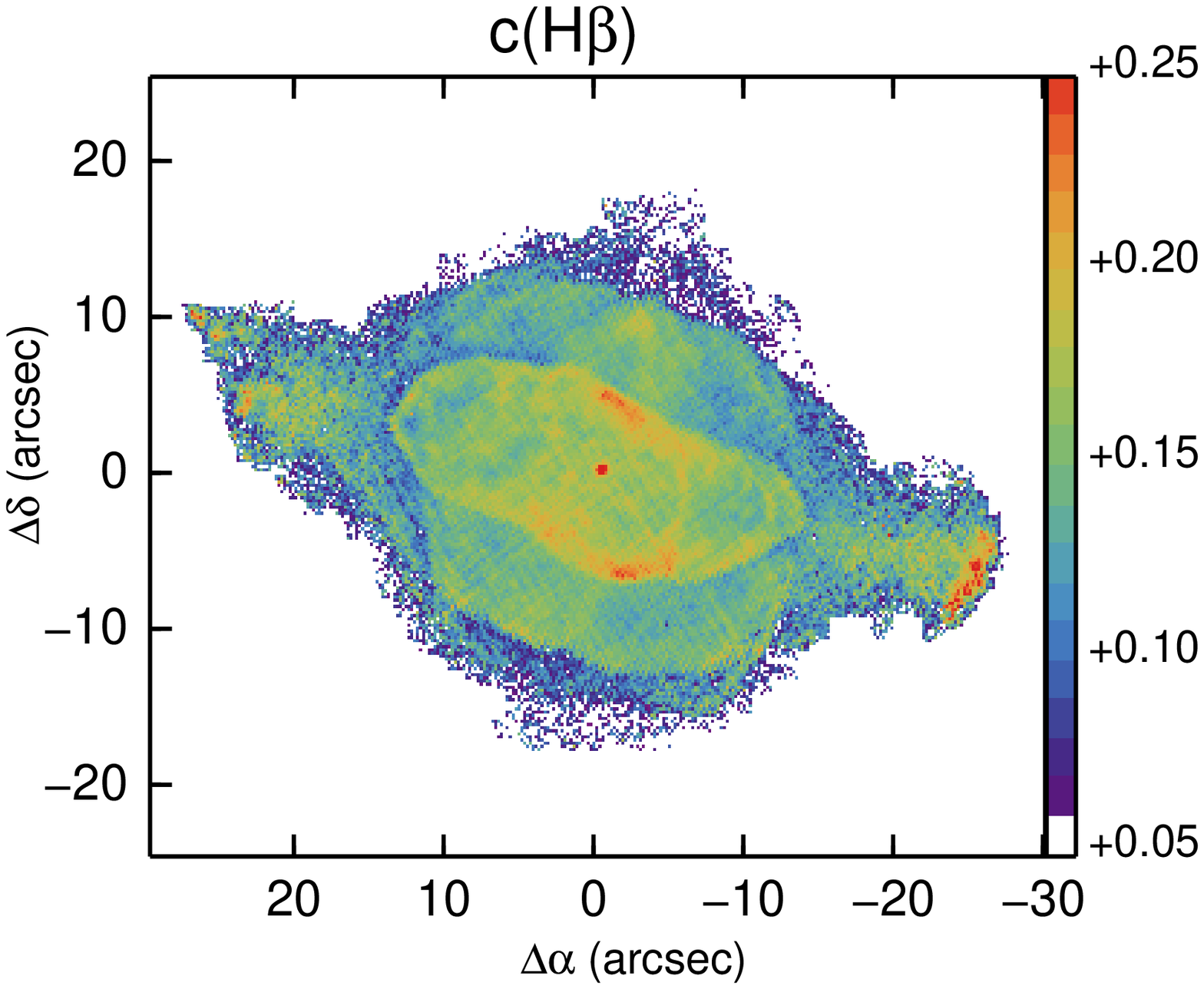}
\includegraphics[angle=0]{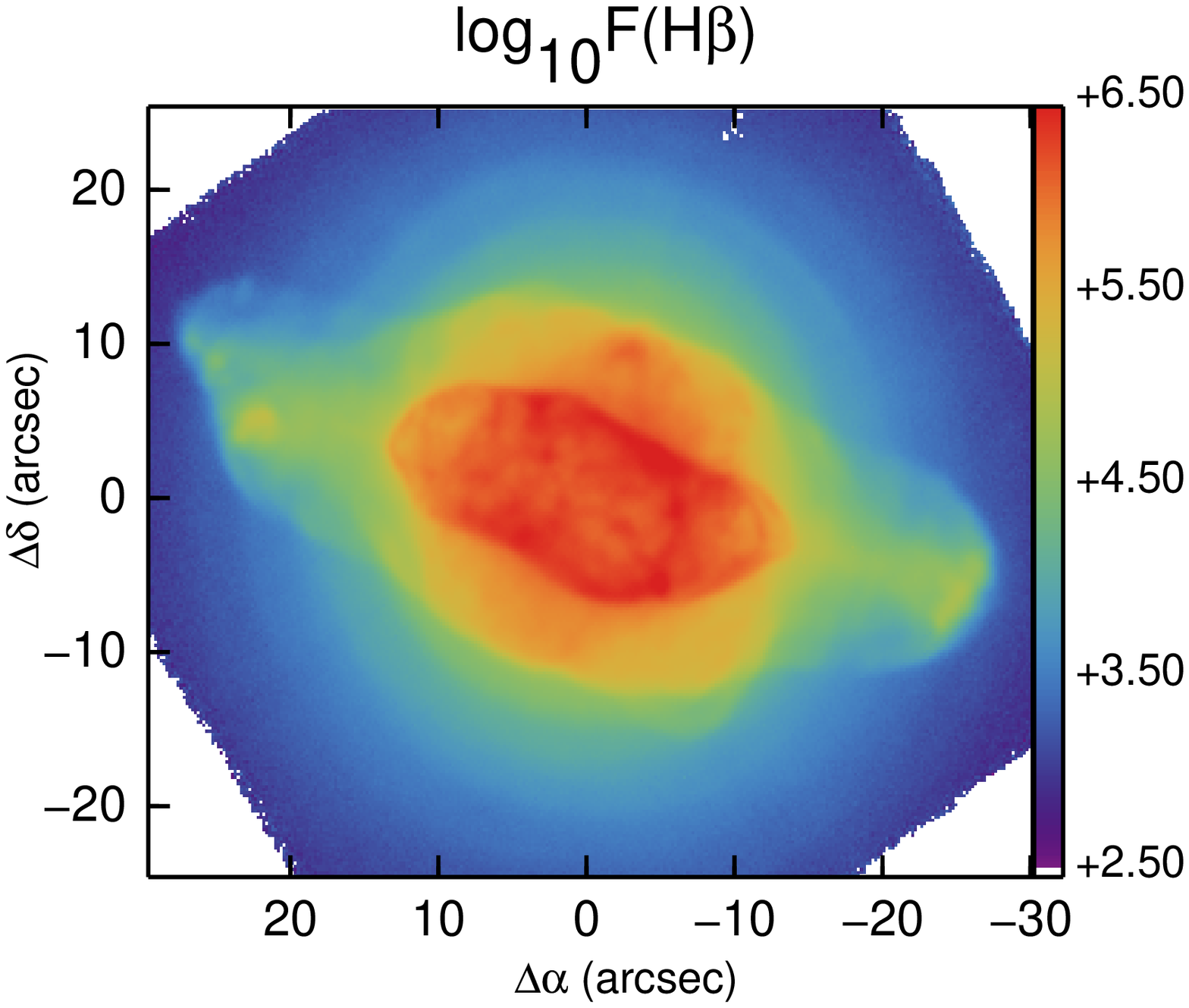}
}
\caption{Left: Log extinction at H$\beta$, $c$, to NGC~7009 based on the 120s H$\beta$ 
and Paschen P9 -- P12 images and the 10s H$\alpha$ flux image, with all images 
corrected for the contamination by He~II and convolved with Gaussians to the 
same FWHM for all images, that of the H$\beta$ image with FWHM 0.56$''$. \newline
Right: Log$_{10}$ H$\beta$ surface brightness image of NGC~7009 based on the 120s MUSE cube.
}
\label{fig:final_cmap}
\end{figure*}

The level of structure in the extinction map (see Fig.~\ref{fig:final_cmap}, 
which is derived from the more careful analysis of the H line ratios described in 
Sects. 3.1 -- 3.4) is surprising, given the low value of $c$ towards this nebula and
the indications of relatively quiescent middle-aged closed shells, apart from the 
more obvious structures such as the caps and the ansae. There is a general
level of concordance between the dust morphology and the emission line / kinematic 
sub-structures of the nebula: the central flat plateau to the inner shell, two regions of
higher extinction to the minor axis inner shell caps, the surrounding lower extinction 
outer shell and several features corresponding to the anse and minor axis polar knots. 
The increased extinction at the position of the central star might not be real but related
to the uncertainty of correcting the H emission line fluxes for underlying stellar 
absorption. The extinction feature at the position of the inner shell-outer shell 
boundary appears as a 'wave', consisting of an inner crest and an outer trough. 
The extinction map of \citet{Tsamis2008} showed a hint of a decrease in $c$ over 
the edge of the inner shell (their Fig. 9), but, since this occurred close to the 
boundary of the IFU area, its reality could not be confirmed. Examination of the 
HST WFPC2 F487N (H$\beta$) and F656N (H$\alpha$) images of NGC~7009, taken in 
Programme 8114, also shows evidence for the wave feature at the edge of the inner 
shell from the simple ratio of the images; a feature not remarked on by 
\citet{Rubin2002}.

Given that this is the first high fidelity map of a PNe taken with MUSE, 
and the ratios of lines at different wavelengths might be expected to reveal artefacts 
due to instrumental effects or data reduction steps, doubts were 
naturally raised that the features in the extinction might be not be real. It is clear from
Fig.~\ref{fig:final_cmap} that some imprint of the image slicer remains in this map as seen 
from the perpendicular striping oriented along the axis of the slicer resulting from
combination of the two views oriented 90$^{\circ}$ apart. This effect could have been 
mitigated by combining one or more cubes with other rotation angles, and clearly several more
exposures at intermediate rotation angles would have been beneficial. However, the principal
features in the extinction map are not correlated with the slicer pattern. 

In order to confirm the reality of the features in this
map, first instrumental, then physical causes were sought to explain these
structures. If these explanations can be countered then the reality of the features
in the extinction map cannot be rejected at this stage. In the following sections,
first instrumental then astrophysical explanations for the apparent structures
in the map are tested.

\subsection{Geometrical distortion with wavelength}
Since the MUSE instrument has some refractive optics (a filter and derotator
in the fore-optics and the spectrograph camera for each integral field unit),
there may be geometrical distortion with wavelength which is not
corrected to the level of precision necessary to register the H emission
line images for careful study of the spatial dependence of the extinction.
Two faint stars at opposite corners of the MUSE field of view were measurable
and, along with the central star, could be used to check for any
uncorrected distortion. One of the faint offset stars was, however, too faint
at the bluest wavelengths ($\sim$4900\AA) to measure reliably. Between
H$\beta$ and H$\alpha$ no distortion could be detected, but over the
range H$\beta$ to Paschen 9 (P9: 9229\AA) a distortion of about 0.6 pixels
was measured from corner-to-corner of the field. Thus
there is a maximum distortion of about 0.2 pixels at the position 
of the wave/trough with respect to the central star between P9 and H$\beta$.
However, the extinction features appear similar from both H$\alpha$/H$\beta$ 
and Paschen/H$\beta$ ratio maps, arguing for a small differential distortion 
with wavelength. By shifting the images, a distortion of $\gtrsim$0.2 
spaxels at the inner rim was found to enhance or weaken the wave/trough feature. 
Given the size scale of 
the major extinction features ($\gtrsim$1$''$), it can be concluded that the 
residual distortion of MUSE with wavelength, over the field of NGC~7009, is 
not critical for this study.

\subsection{Saturation / detector non-linearity}
The e2V Deep Depletion CCD's of the MUSE detectors have a tabulated saturation
of 65000e$^{-}$. While the H$\beta$ line is not saturated in any of the exposures, 
the H$\alpha$ line is saturated in the 60s and 120s exposures as evident by the 
presence of zeros in the peak of the line in the spectra of the brightest spaxels,
which occur inside the main shell. The transition inner-outer shell, where the 
prominent extinction feature is evident, occurs at lower H$\alpha$ surface brightness 
than towards the centre of the inner shell and so is not clearly affected by 
saturation. However, non-linearity may be occurring in this vicinity to affect the 
H$\alpha$ line peak, perhaps causing an effect on the measured 
extinction. The H$\alpha$ surface brightness changes by about a factor two 
across the inner-outer shell and any non-linearity in the brighter inner shell 
emission would depress the H$\alpha$ line count causing lower extinction to be 
recorded, whilst Fig.~\ref{fig:final_cmap} shows that $c$ is elevated just inside 
the shell, and lower outside where the H$\alpha$ 
flux is lower. It appears fairly convincing that the wave-trough behaviour 
of the extinction across the rim of the inner shell is not due to detector 
saturation/non-linearity and the observation of the same pattern to the extinction 
measured from weaker H lines (see below) confirms this conclusion.

\subsection{Wavelength dependent seeing mismatch}
The atmospheric seeing varies with wavelength, approximately as $\lambda^{-1/5}$ 
\citep{Fried1966} and the effect on measured image quality was seen by the smaller 
FWHM of star images at redder wavelengths compared to 4900\AA\ in the MUSE cubes. 
Given that the wave feature of 
the extinction maps occurs where the surface brightness decreases by a factor 
$\sim$2 over a distance of a few spaxels ($<$0.5$''$), then the effect of the different 
image FWHM between the wavelengths of the H Balmer lines would modulate the ratio image
which forms the basis of the extinction determination. 
For example for a step function across the shell, the ratio map H$\alpha$/H$\beta$ 
would be expected to show a trough, since the image FWHM at H$\alpha$ is lower than at 
H$\beta$. On the 10s MUSE cube the measured FWHM of the central star varies from 0.53 
to 0.50$''$ between H$\beta$ and H$\alpha$. In order to determine if this effect could
be responsible for the observed wave-trough appearance across the inner shell,
the H$\alpha$ image was convolved with a Gaussian to the same measured FWHM as
the H$\beta$ image and the extinction map redetermined. The strength of the trough is
slightly decreased, by $<$0.02 in c, and the wave increased by a lesser amount,
but the features remain. A repeat of this test but with the 120s H$\beta$ map
showed a similar result.

Having rejected instrumental or seeing effects as causing the pattern in 
extinction, astrophysical effects are now examined.

\subsection{Balmer {\it v.} Paschen line ratios}
The He~II Pickering series 6g-4f 6560.10\AA\ line is within 2.7\AA\ of the H$\alpha$ 
line and at the resolution of the MUSE spectra, and given the very large flux ratio 
($\sim$100), will be included in the flux of the fitted H$\alpha$ line. Thus the 
H$\alpha$ flux needs to be corrected for the contamination of the He~II line
($\sim$1\% of the H$\alpha$ surface brightness in the bright He~II core of the nebula). The 
He~II Pickering 7g-4f 5411.52\AA\ line was measured in all the cubes and from the 
Case B ratio 6560.1/5411.5\AA\ ratio, the 6560.1\AA\ observed line flux was 
computed using \citet{Storey1995} and a reddening map. The observed value of 
the He~II 6560.10\AA\ line of 0.0338$\times$F(H$\beta$) tabulated by \citet{Fang2011}
appears to be too large (the theoretical ratio 6560.1/5411.5\AA\ is 1.75 at 
10$^{4}$K whilst \citet{Fang2011} have a dereddened ratio of 2.92). Similarly the
Pickering He~II 8g-4f 4859.32\AA\ line is blended with H$\beta$ at the MUSE resolution,
so its contamination of the measured H$\beta$ flux should also be corrected
(similarly $\sim$1\% of H$\beta$ in the core). The theoretical ratio of 
4859.3\AA\ to 5411.5\AA\ was again used. 

Applying both these corrections to the H$\alpha$ and H$\beta$ flux maps, the 
derived extinction map was re-computed. The He~II line is indeed only strong inside
the inner rim and is very weakly present outside the rim, so the effect of the
correction to H$\alpha$ and H$\beta$ fluxes for He~II is insignificant where the 
strongest features of the extinction are present. Inside the inner rim, where the
He~II is brightest, the effect on the extinction map is small, c$<$0.01.

The Paschen lines higher than P9 (9229.0\AA) are also available for extinction
determination. These lines also have the advantage that the contaminating 
He~II lines (from n=6) are weak, $<$0.002$\times$ 4f-3d 4685.68\AA\ line (or
$<$0.02 $\times$ He~II 7g-4f 5411.52\AA\ line) but still a fraction $\sim$0.02 of 
the nearby H I line. The Paschen lines were thus also corrected for the contamination
by the nearby He II n=6 line. The logarithmic extinction correction at H$\beta$ can then be
determined from H$\beta$, H$\alpha$, and as many lines as desired above P9. It was
decided not to use Paschen lines above P12 (8750\AA) as they become faint ($<$1\% of 
flux of H$\beta$), more crowded and lose the wavelength leverage relative 
to H$\beta$.

\begin{figure}
\centering
\resizebox{\hsize}{!}{
\includegraphics[angle=0,clip]{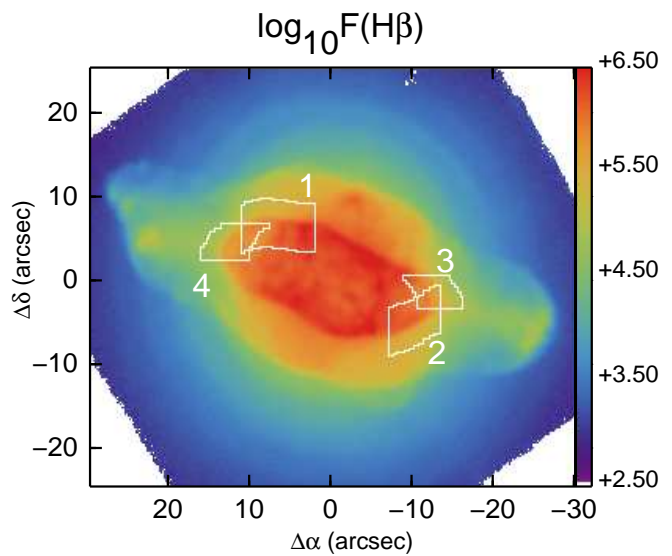}
}
\caption{Log$_{10}$ F(H$\beta$) image of NGC~7009 from the 120s cube with
the positions indicated for the four extraction regions across the bright rim
(see Fig~\ref{fig:rimsectc}).
% Regions numbered as 3 and 4 should be reversed. Done in masklgHbflux.ps
}
\label{fig:rimregions}
\end{figure}

\subsection{Temperature and/or density structures}
The ratios of the H Balmer lines and the Paschen to Balmer lines are dependent on
both $T_{\rm e}$ and $N_{\rm e}$, so the extinction map from the Balmer and Paschen lines
should use both the $T_{\rm e}$ and $N_{\rm e}$ maps to determine the intrinsic H 
line ratios. The $T_{\rm e}$ map was determined from the [S~III]9069/6312\AA\ ratio 
map and the $N_{\rm e}$ map from the [Cl~III]5518/5538\AA\ ratio (to be 
presented in Paper II). Since the line ratios are affected by the reddening, 
they require to be reddening corrected before $T_{\rm e}$ or $N_{\rm e}$
are determined. Initially a constant extinction value of $c$=0.15 was employed, $T_{\rm e}$ 
and $N_{\rm e}$ calculated from the dereddened [S~III] and [Cl~III] ratios, respectively, 
and then these maps used to recalculate the intrinsic H line ratios. It was found that 
another iteration of this procedure (also including $T_{\rm e}$ or $N_{\rm e}$ maps determined
from dereddened line ratios) had no effect on the extinction values exceeding the 
errors. Fig.~\ref{fig:final_cmap} shows the resulting $c$ map based on the 120s H$\beta$ 
and Paschen P9 - P12 images and the 10s H$\alpha$ flux image, corrected to the same FWHM 
for all images, as described in sect.~3.2, and contamination by He~II as described in 
Sect.~3.3. The map displays mean $c$ from the observed to theoretical H line ratios
(H$\alpha$/H$\beta$, P9/H$\beta$, P10/H$\beta$, P11/H$\beta$ and P12/H$\beta$) weighted by
the inverse square of the errors determined from the observed line ratios.

An obvious cause of the wave over the inner rim could be temperature and/or density changes. 
The $T_{\rm e}$ map from [S~III]9069/6312\AA\ shows evidence for
a mild temperature profile across the rim, with $T_{\rm e}$ elevated by $\sim$200K over 
the $c$ trough and slightly depressed in the $c$ wave (see for example the 1D profile 
in Fig.~\ref{fig:rimsect4}). The [S~III] emission line 
images used to determine $T_{\rm e}$ were reddening corrected by the 
extinction map, but if an extinction map with a constant value is used to
deredden the [S~III] lines, the $T_{\rm e}$ profile is more enhanced 
($\sim$300K in the $c$ trough).

\subsection{Investigation of $c$ across the inner shell}
\begin{figure*}
\centering
\resizebox{\hsize}{!}{
\includegraphics[width=0.97\textwidth,angle=0, bb=0 170 530 380,clip]{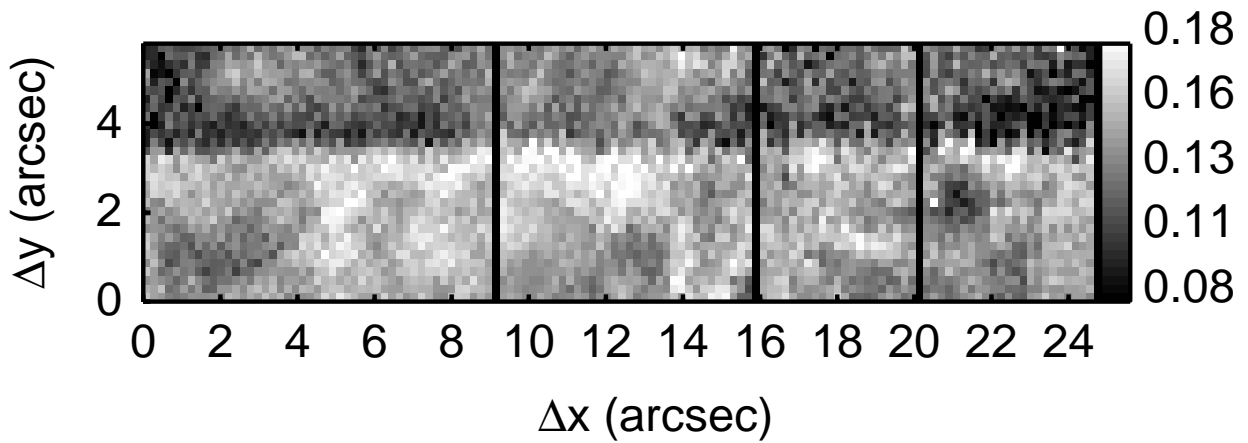}
}
\caption{Images of log extinction at H$\beta$, $c$, over four cross section regions of 
the inner shell. See Fig.\ref{fig:rimregions} for the positions of the four sections.
Sections 1-4 are displayed left-right, with a one pixel gap between each section.
The direction towards the central star is downwards.
}
\label{fig:rimsectc}
\end{figure*}

In order to display the behaviour of $c$ and related quantities,
a number of sections across the inner rim were assembled as representative of the
feature. Profiles along the cube $\alpha$,$\delta$ pixel grid were constructed in four regions 
at the north and south edges of the inner shell and the SW and NE extremities of the
shell major axis; Fig.~\ref{fig:rimregions} shows the four regions on the 
log H$\beta$ image. Without resampling, the individual pixel profiles were assembled 
into an image aligning them along the rim. Fig \ref{fig:rimsectc} shows the resulting 
image for the extinction. Collapsing this image, or part of it, perpendicular to 
the rim, enables a detailed consideration of the overall behaviour of
this feature.

\begin{figure*}
\centering
\resizebox{\hsize}{!}{
\includegraphics[angle=-90]{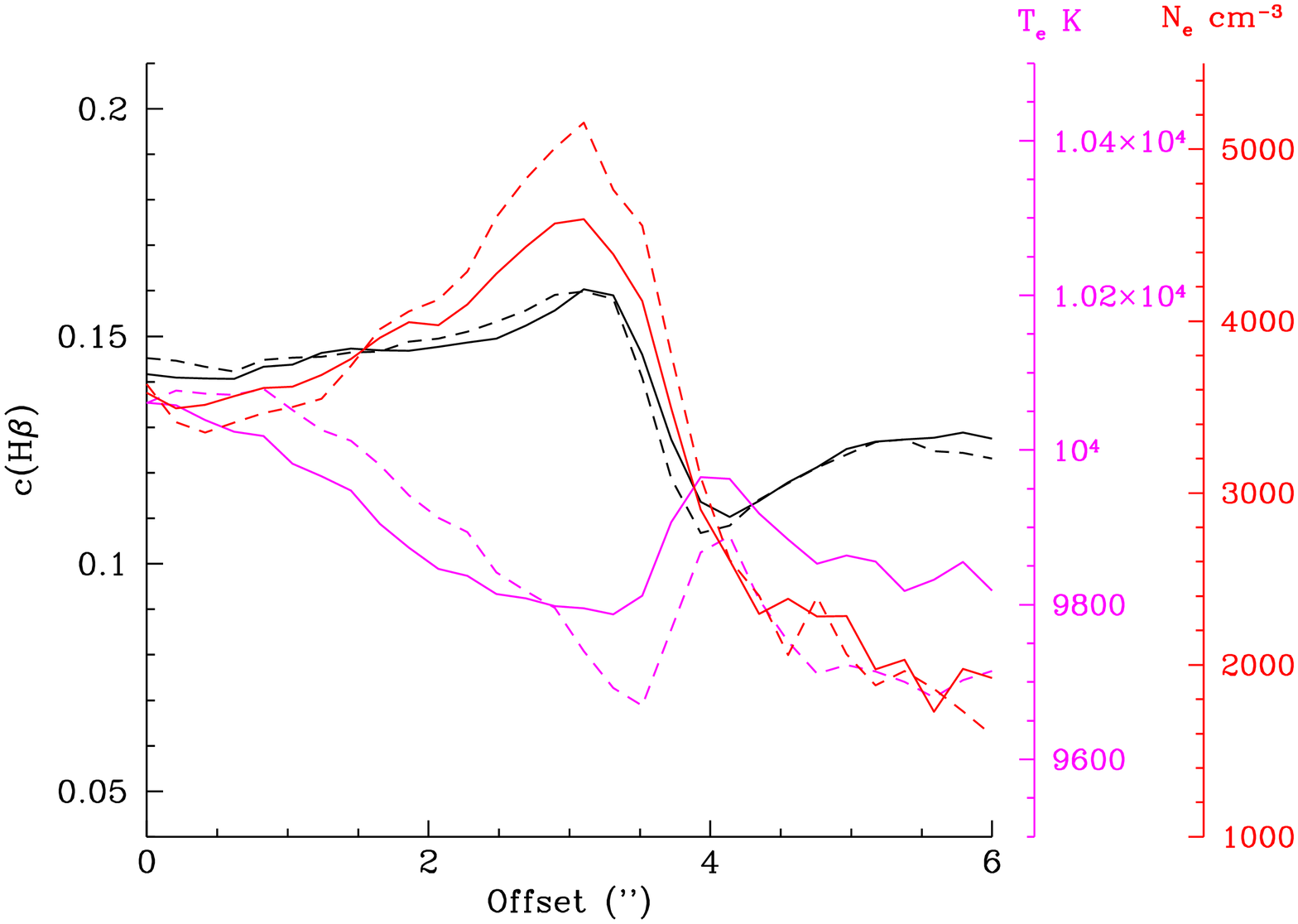}
\includegraphics[angle=-90]{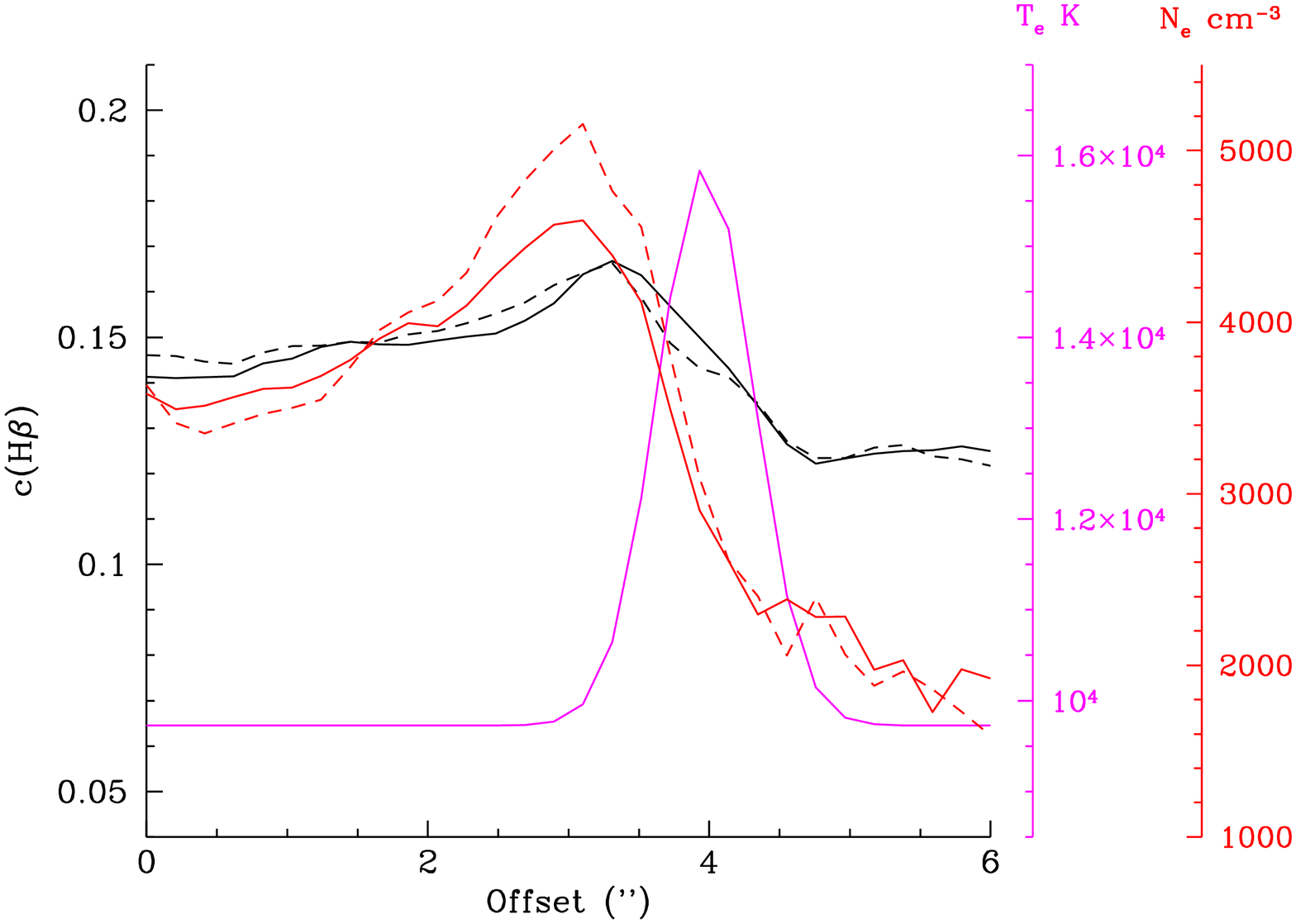}
}
\caption{Collapsed cross sections for the four rim regions displayed in 
Fig.~\ref{fig:rimregions} (a total extent of 25$''$), showing the variation of
extinction, $c$, $T_{\rm e}$ (K) and $N_{\rm e}$ (cm$^{-3}$) with relative offset across
the rim (solid lines). The dashed lines refer to the mean values across the 
rim region 1, to the NE. \newline
Left: cross sections for $N_{\rm e}$ from the mean value determined from
the [S~II]6716/6731\AA\ and [Cl~III]5518/5538\AA\ ratios and $T_{\rm e}$ from the
[S~III]9069/6312\AA\ ratio and theoretical Balmer and Paschen line ratios 
computed for these values. \newline
Right: instead of measured $T_{\rm e}$, an impulse function (Gaussian with FWHM 0.9$''$)
is used with the observed $N_{\rm e}$ to compute $c$ from the H line ratios. The size of
the temperature excursion (6000K) was chosen to remove the trough in the extinction.
}
\label{fig:rimsect4}
\end{figure*}

The average behaviour of the extinction, $c$, across the rim in the direction perpendicular 
to the rim, with offset increasing with distance from the central area of the nebula, is 
shown in Fig.~\ref{fig:rimsect4}. On this plot the mean value of $c$ across the
four regions is shown (bold line) together with the profile for the region to the NE
which displays the trough in extinction most strongly (dashed line). Similarly,
profiles of $N_{\rm e}$ and $T_{\rm e}$ are shown for the same regions. The left plot 
shows the $N_{\rm e}$ profile as formed from the mean $N_{\rm e}$ derived from the 
[S~II]6716/6731\AA\ and [Cl~III]5518/5538\AA\ ratios (to be presented in Paper II). The
profiles of $c$ in Fig.~\ref{fig:rimsect4} are formed from the mean across the regions
shown in Fig.~\ref{fig:rimregions}, similarly for $N_{\rm e}$ and $T_{\rm e}$ (i.e., surface 
averaged profiles are displayed).

To determine if changes in $T_{\rm e}$, or $N_{\rm e}$, across the rim could be 
responsible for the behaviour of the extinction, an impulse profile, centred
on the minimum of the trough, was synthesised to simulate rapidly varying $T_{\rm e}$, 
or $N_{\rm e}$. Considering $T_{\rm e}$, a positive excursion of temperature by about 
6000K was applied: recomputing the H line theoretical strengths could then remove 
most of the trough, as shown in Fig.~\ref{fig:rimsect4}, right hand plot. No clear evidence 
for such large changes in $T_{\rm e}$ have been seen in previous observations (e.g.,
\citet{Rubin2002} and \citet{Tsamis2008}), nor in the MUSE [S~III] $T_{\rm e}$ map 
(to be presented in Paper II).

A similar approach with an impulse for $N_{\rm e}$ produced only a marginal difference in 
the extinction profile (slightly deeper $c$ drop at 4$''$ offset for density 10$^{6}$ 
cm$^{-3}$ at the peak), since the Balmer and Balmer-to-Paschen line ratios vary only 
slightly with densities over the range 1 cm$^{-3}$ to 1$\times$10$^{6}$ cm$^{-3}$.

\section{Discussion}
\subsection{Interstellar reddening}
Given that the interstellar extinction in the direction towards NGC~7009 ($l,b$ 37$^\circ$.8, 
-34$^\circ$.6) tabulated by \citet{Schlafly2011} is greater than the low extinction values mapped, 
then NGC~7009 must lie at an intermediate distance along the extinction column derived
by \citet{Schlegel1998} from the 100$\mu$m emission maps. The low values of extinction,
over the well-measured region of bright emission, range upwards from $c$=0.04 with typical 
errors of 0.01. Over a larger area including the ansae and faint regions on the minor
axis, the fraction of spaxels with non-zero extinction rapidly increases above $c$=0.04 
(some spaxels with lower extinction values occur, but can include this value within 
errors). Thus a likely value of extinction between the observer and the nebula
is 0.04, and subtraction of this value from the measured extinction to NGC~7009 results
in a map of the intrinsic extinction in the intra-nebular and circum-nebular volume.

This estimate of the line-of-sight ISM extinction to NGC~7009 can be compared to the 
three dimensional maps of dust extinction in the Milky Way from Pan-STARRS 1 and 2MASS 
photometry by \citet{Green2015}. For a $c$=0.04 ($E_{B-V}$ 0.028 mag.) ISM contribution, 
the resulting distance to the nebula is only 270 pc. However, the error range of 
$E_{B-V}$  {\it v.} distance modulus in the direction to NGC~7009 encompasses 0 and 
$>$1000 pc, thus not incompatible with the estimated distance of 1400 pc 
\citep{Sabbadin2004}, in good agreement with the value of 1450 pc deduced by 
\citet{Frew}.

\subsection{Dust-to-gas ratio}
With the map of the intrinsic extinction of the nebula (from Fig.~\ref{fig:final_cmap}), 
and given the distribution of the nebular line-of-sight column of H, $N_{H}$, 
the dust-to-gas ratio (D/G) can be determined. There are two methods to determine
the line of sight H column: from the measured $N_{\rm e}$ and an estimate 
of the line of sight depth of the nebula; or from the emission of H which provides
$ \int N_{H} \times N_{e} dl$ along the line of sight. The former method
involves a guess on the line of sight emitting depth of the nebula together with a
correction for the number of electrons per H atom. The latter method does not require an 
assumption of line-of-sight depth since the integral of the H column density is 
determined from the surface brightness in an H line and the line emissivity; the depth  
variation of electron density is not known, so the assumption of a constant value
along the line of sight is adopted. The former method requires the distance for the 
estimate of the line of sight column, and so the latter method, not requiring this 
assumption, is preferred.

\begin{figure}
\centering
\resizebox{\hsize}{!}{
\includegraphics[angle=0]{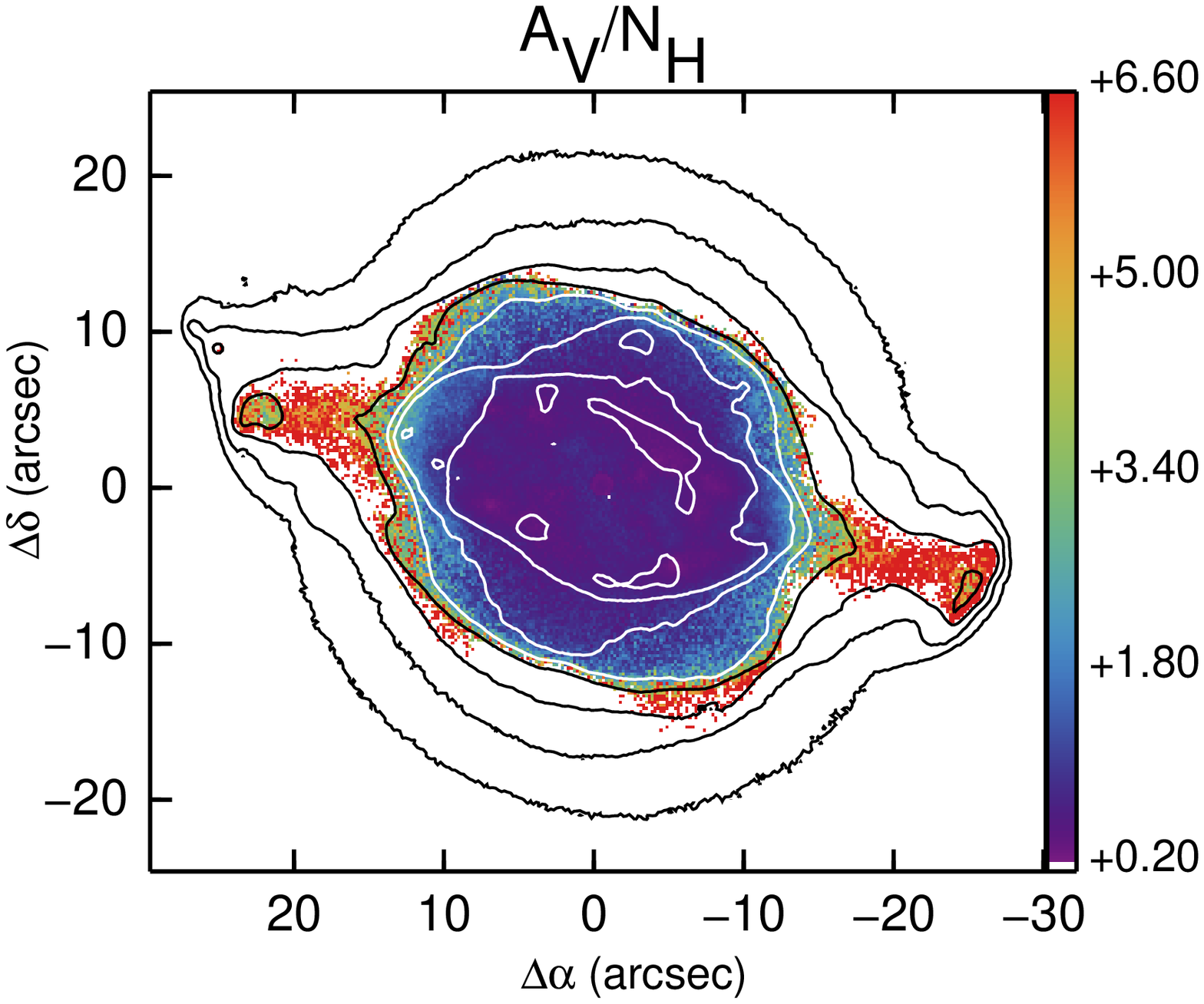}
}
\caption{The dust-to-gas ratio map of NGC~7009 as expressed by the ratio 
$A_V/N_{H}$ in terms of the MW ISM value of \citet{Gudennavar2012}. The 
contours correspond to the log$_{10}$ F(H$\beta$) image of NGC~7009 
from the 120s cube, shown in Fig.~\ref{fig:final_cmap}.
}
\label{fig:dust_gas}
\end{figure}

\begin{figure}
\centering
\resizebox{\hsize}{!}{
\includegraphics[angle=-90]{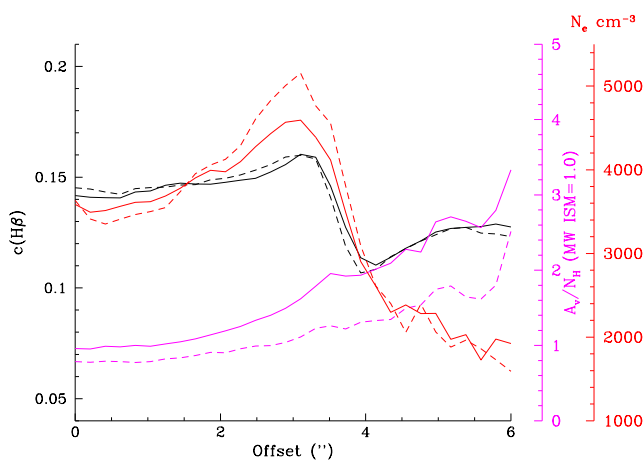}
}
\caption{Collapsed mean cross sections for the four rim regions displayed in 
Fig.~\ref{fig:rimregions}, showing the variation of extinction, c, $A_V/N_{H}$ 
and $N_{\rm e}$ (cm$^{-3}$) with relative offset across the rim (solid lines), and 
for rim region 1 (dashed lines); cf., Fig.\ref{fig:rimsect4} . The dust-to-gas
ratio $A_V/N_{H}$ is 
expressed as a fraction of the ISM value of $5.07 \times 10^{-22}$ mag. cm$^{-2}$ 
from \citet{Gudennavar2012}.
}
\label{fig:rimsectNh}
\end{figure}

From the H$\beta$ flux map, with the [Cl~III]5518/5538\AA\ electron density and
the electron temperature from [S~III]9069/6312\AA\ to calculate the H$\beta$ emissivity 
\citep{Storey1995}, $N_{H+}$ was computed. Then ratioing this with the extinction map,
after subtracting the interstellar value, results in a map of $A_V/N_{H}$.
The presence of only weak lines of neutral species, such as [O~I],
which will be presented in Paper II, confirm the small contribution of neutral H 
to the total H budget, so that $N_{H+} = N_{H}$ can be assumed. 
For convenience the $A_V/N_{H}$ map was expressed as a fraction of the Milky Way
ISM value of 5.07 $\times 10^{-22}$ mag. cm$^{-2}$, from the mean $N_{\rm H}/E_{B-V}$ value 
quoted by \citet{Gudennavar2012} (adopting R=3.1 since $N_{\rm H}/E_{B-V}$ is quoted) and the
resulting map is shown in Fig.~\ref{fig:dust_gas}. Over the inner shell, the mean value 
is about 0.9, but rises in the outer shell and the streams to values around 5 (as in this 
region the surface brightness in H lines is lower, the extinction slightly depressed relative 
to the inner shell and the density about a factor 2 lower). The $A_V/N_{\rm H}$ profiles 
across the rim sections are plotted in Fig.~\ref{fig:rimsectNh} similarly to Fig. 
\ref{fig:rimsect4}, including also the profiles of $c$ and $N_{\rm e}$.

The D/G map (Fig.~\ref{fig:dust_gas}) can be compared with the map produced
from $N_{H}$ using the line-of-sight depth and $N_{\rm e}$. A line-of-sight depth of
20$''$ was used (based on an prolate nebula form, cf., \cite{Steffen2009}, and see Sect.
\ref{sect:Model}) and a ratio $N_{\rm e}/N_{\rm H}$ of 1.14 (based on the chemical 
abundances listed by \citet{Sabbadin2004}). The pattern of $A_V/N_{H}$ is similar 
(not shown), in particular the rising value in the outer regions. The absolute 
values are smaller by a factor $\sim$2, suggesting a filling factor value of around 
50\% between the H density and the matter density based on $N_{\rm e}$, or a distance 
of 750 pc for an assumed filling factor of 1.0.

The integrated $A_V/N_{H}$ for NGC~7009 with respect to the interstellar value can 
be computed in several ways depending on how the surface weighting is performed. 
Weighting the $A_V/N_{H}$ map by the observed H$\beta$ flux map, results in an integrated
$A_V/N_{H}$ value of 0.87; if weighted by $ \int N_{H+} dl$, the value is 0.99; 
alternatively weighting by $\sqrt{H\beta flux}$ gives a value of 1.38; and finally 
weighting by the electron density map (mean of $N_{\rm e}$ from [S~II] and [Cl~III] 
ratios) gives an integrated $A_V/N_{H}$ value of 2.21. Weighting by electron density 
is the least comparable to the ISM value determined from studies of the extinction 
and neutral gas column to nearby stars; the value integrated by the PN $ \int N_{H+} dl$ 
is the most comparable, but for unresolved PNe, the weighting will, by default, be by 
H$\beta$ flux. 

The integrated $A_V/N_{H}$ value can be compared with the ionized gas-to-dust mass 
ratio derived from the spectral energy distribution fit, including the millimetre 
and sub-millimetre measurements from Planck, of 235 \citep{Planck2015}, although 
with an uncertainty of $\pm$50\%. The line-of-sight dust extinction is calculated 
assuming that the dust absorbs and scatters the emission from the gas. With the 
assumption that the dust is uniformly mixed with the ionized gas in proportion to 
the density, then the measured line-of sight dust extinction is about a factor two 
less than the total amount of dust available for extinction (i.e., if all the dust 
was present in the foreground), depending on the variation of emission 
($\propto$ density$^{2}$) with radius. Correcting for the mass
fraction of metals (35\%, from \citet{Sabbadin2004}) and assuming astronomical
silicate grains \citep{Draine1985}, an estimate of the mass of the dust responsible 
for the extinction can be made from $C_{ext}$ calculated from Mie theory and the 
mass density ($\rho_{Sil}$ = 3.2 g cm$^{-3}$; \citet{DraineLee1984}). 

% New check
% /home/jwalsh/mie/makeD_G2.f
% Assuming Av for dust mixed with gas MRN 0.005,0.25 extincting dust only accounts 
% for 85% of the mass for G/D=235 and integrated Av/NH = 0.99 / 0.45 
% For 235 => MRN Si 0.007,0.25 for 100% match extinction = emission masses. 
% For 235 - 50% = 117.5 => MRN Si 0.035,0.25
% For 235 + 50% = 353.0 => MRN Si 0.003,0.25
%

Assuming an MRN \citep{Mathis1977} spherical grain size distribution for silicate 
(size distribution $\propto a^{-3.5}$ and $a_{min}$ 0.005 $\mu$m, $a_{max}$ 0.25 $\mu$m), 
the mass of obscuring dust is only responsible for about 85\% of the dust mass 
from thermal emission. If the lower bound on the grain radius is increased from the 
standard MRN silicate value, to 0.007$\mu$m, then the masses of obscuring and 
emitting dust can be made equal. The upper bound for the 50\% error on the 
\citet{Planck2015} gas-to-dust mass ratio is compatible with the $A_V/N_{H}$ value 
for the standard MRN silicate ISM size distribution. A greater mass in thermal 
dust, from cooler or warmer dust components than measured from the SED fits, 
would increase the mismatch with the extinction dust mass, requiring the presence 
of larger grains. 
% On the other hand, non-spherical grains have a lower extinction per unit mass 
% which could permit compability with the dust in NGC~7009 and 'standard' ISM dust.

It is noteworthy that both (i) the integrated $A_V/N_{H}$ (based on 
$ \int N_{H+} dl$ weighting and corrected for the effects of extinction 
measurement from gas mixed with dust using the model presented in
Sect. 4.4) of 2.2 relative to the ISM value and (ii) the 
integrated gas-to-dust mass ratio of 235 from \citep{Planck2015}, are  
similar to the ISM values (256 for the latter using the determination by
\citet{Draine2011} and He/H=0.1). The 
conclusion is that NGC~7009 is marginally a net dust producer. Since the 
$A_V/N_{H}$ measurements only pertain to the dust in proximity to the ionized gas, 
such a conclusion would be modified by the presence of a more extended dust (and gas) 
component which is the remnant of the long-term mass loss on the AGB before the PN phase. 
NGC~7009 does not, however, have an enhanced C/O ratio, and its O/H value is slightly
below solar, so its heavy element fraction is probably similar to that of the 
ISM out of which its progenitor star formed. The fraction of its heavy elements 
that form dust during the AGB phase is thus likely to be similar 
to the original ISM fraction, since the CNO cycle only recycles elements already 
present in the progenitor star.

However, PNe which have C/O$>$1, as well as Type I PNe whose CNO/H ratio
significantly exceeds that of the ISM, have undergone the triple-alpha process, 
producing primary carbon (some of which may later have been converted to N by 
hot bottom burning). Such PNe may have $A_V/N_{H}$ ratios that considerably 
exceed the ISM value and similar MUSE observations to those on NGC~7009 could 
be indicative.

\subsection{Effect of $T_{\rm e}$ variation}
If the ionic temperature is much higher across the inner shell boundary (explored by 
applying a $T_e$ impulse across the extinction trough), whilst showing only
small evidence of raised temperature in the [S~III] $T_e$ (see Fig.~\ref{fig:rimsect4}),
then it is suggested that the lower $c$ domain outside the bright rim might also be 
attributable to higher H temperatures. This would, however, imply lack of equipartition 
between electrons and H$^{+}$. Interestingly the PN hydrodynamic model A4 of
\citet{Perinotto1998} does show an $\sim$5000K temperature rise at the outer rim 
for an optically thin nebula, corresponding to the position of the isothermal shock.

From studies of the H temperature from the Balmer (and Paschen) jump in PNe 
(\citet{Zhang2004}; \citet{Liu1993}), it was found that 
$T_{\rm e}$(Bal) is generally systematically lower than $T_{\rm e}$ determined 
from [O~III] 5007/4363\AA. In the case of NGC~7009, the Balmer and Paschen jump 
temperatures were determined as 7200K and 5800K, respectively \citep{Zhang2004}; thus 
3000-4000K lower than from the forbidden line temperatures ([O~III] and [S~III]). 
The region sampled by the spectra used to fit the Balmer and Paschen jumps is not 
explicitly specified, but appears to correspond to the bright main shell. Although 
it cannot be stated with certainty that there is no temperature impulse at the
main shell rim applicable to H and coincident with the extinction trough, this 
appears unlikely given that the majority of PNe (cf., \citet{Zhang2004}) show 
T$_e$(Bal) $<$ $T_{\rm e}$[O~III].

\subsection{Possible model}
\label{sect:Model}
In order to better understand the morphology of the extinction and $A_V/N_{H}$, a very simple
3D geometrical model of the emission and dust was constructed. Two co-centred ellipsoidal 
forms for the gas emission and dust density were chosen as the minimum required to
model the gross features; based on the $Shape$ model of \citet{Steffen2009}, prolate
ellipsoids were used. The projected emission model was guided by the H$\beta$ image 
(Fig.~\ref{fig:final_cmap}) and the two ellipsoids, or ellipsoidal shells, had
differing radial emission dependence or shell thickness, with sizes approximately
scaled to the relative dimensions of NGC~7009. Ellipsoidal shells were used for
the dust number density but with differing radial dependence. The 3D models were
projected along the line of sight, without considering any body tilt of the ellipsoids. 
Assuming an extinction coefficient appropriate for small silicate grains, the model 
emission is extinction corrected depending on the length of the line-of-sight dust column 
between the emitting point and the observer. Projected maps of the emission, with and
without internal extinction, are output and the $c$ map is calculated from their
ratio by analogy to the observed map (Fig.~\ref{fig:final_cmap}), which used the observed 
images of several H lines to determine $c$. 

\begin{figure*}
\centering
\resizebox{\hsize}{!}{
\includegraphics[angle=-90]{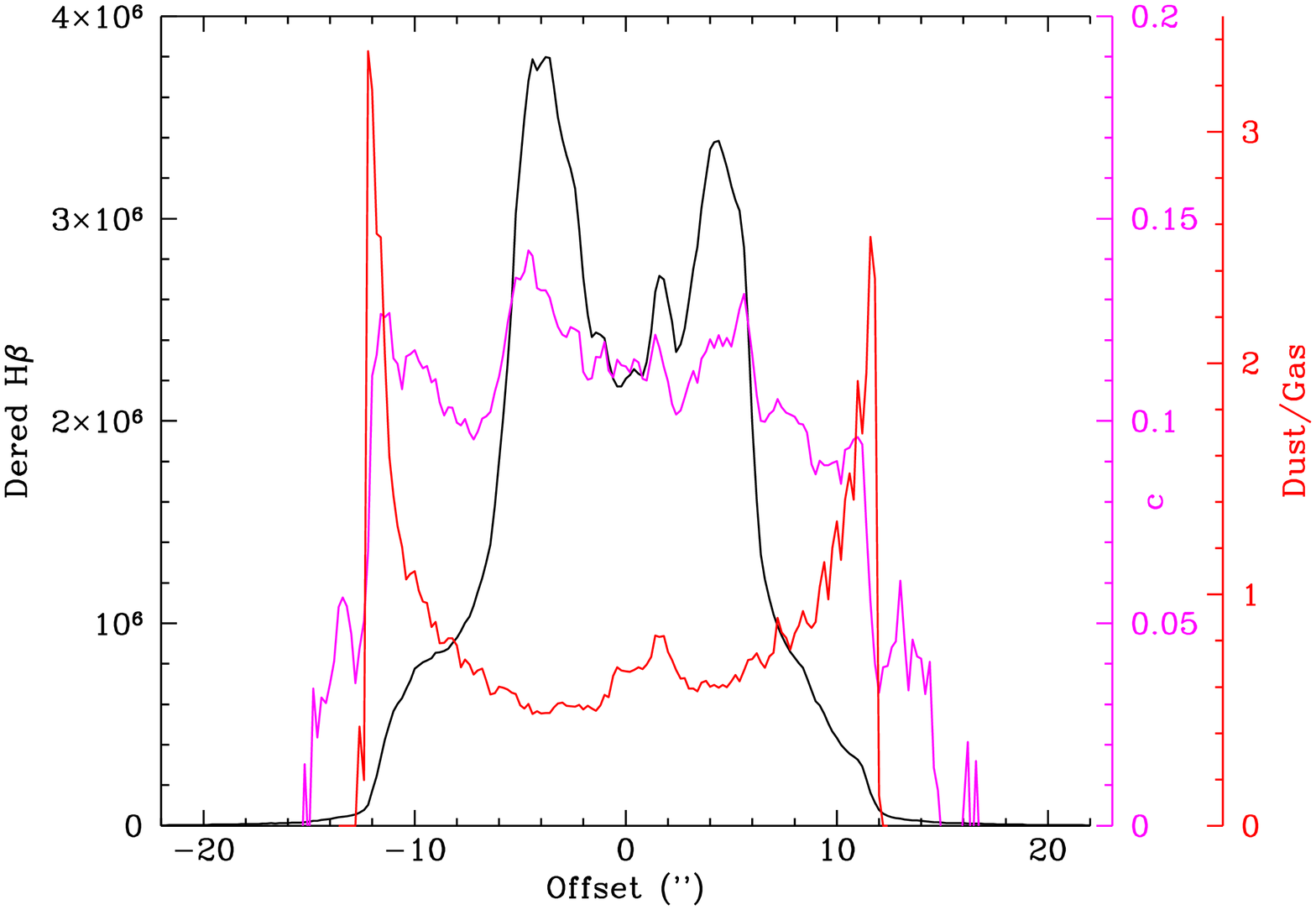}
\includegraphics[angle=-90]{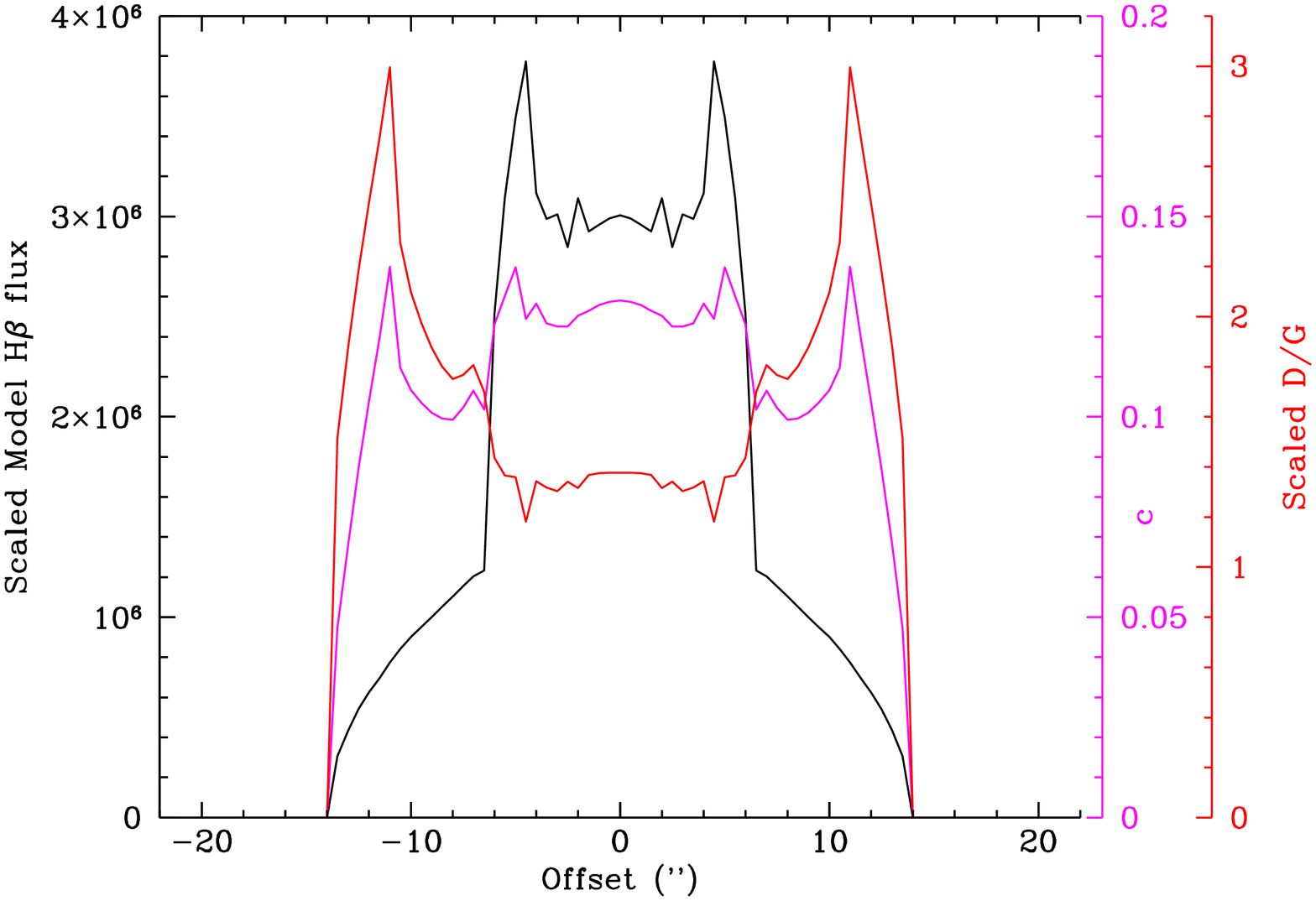}
}
\caption{Left: The observed profiles of the H$\beta$ flux (relative units, in black), 
extinction $c$ (magenta) and dust-to-gas ratio ($A_V/N_{H}$, in red) along a strip 2$''$ wide 
along the minor axis of NGC~7009. The extinction refers to the nebular component and
an assumed interstellar component (of $c$ = 0.04) has been subtracted. \newline
Right: Plausible model profiles of emission (assumed H$\beta$), extinction $c$ and
dust-to-gas ratio (from sums of model gas and dust columns) plotted with scaling and plot
symbols similar to the observed profiles. The model consists of two prolate ellipsoidal
dust shells; see text for details. 
}
\label{fig:Obs_Mod}
\end{figure*}

Various radial dependencies for emission and dust in the two ellipsoids representing the
inner and outer shells were tried. Figure \ref{fig:Obs_Mod} (left) shows a slice along the 
minor axis of the nebula taken 3.5$''$ east of the central star for the H$\beta$ flux, 
extinction $c$ and $A_V/N_{H}$. The minor axis was selected since the ansae and FLIERS 
occur on the major axis. These profiles were qualitatively compared to the model 
profiles at the equivalent position (not a fit), with model D/G taken as the ratio of dust 
density to model density ($\propto \sqrt{flux}$). A fair match was found with a constant 
density inner emission ellipsoidal shell, and a filled volume with $r^{-0.5}$ for the 
outer region; for the inner dust ellipsoid, dust density $\propto r^{-0.3}$ was adopted and 
$r^{-2.5}$ for the outer dust shell. Figure \ref{fig:Obs_Mod} (right) shows the model
profiles, with the emission and D/G arbitrarily scaled to the same range as the observations.
The normalization for the extinction is provided by the dust density in the model. While
physical parameters cannot be directly determined from these models, which are by no means 
unique, a few features are obvious: a prolate nebula does not conflict with the gas and
dust imaging; the radial dependence of the emission differs in the 
inner and outer shells; the dust and emission (gas) radial dependencies differ (in order to 
give D/G increasing with distance from the nebular centre); at least two dust shells are 
necessary to explain the extinction structure; the outer volume is more filled by dust 
(shell thick) to provide the extinction in front of the inner shell.  

The main aim of these simple models was to explore if a combination of different emission
and dust shells could give rise to the trough in the extinction at the outer edge of the
inner shell. The combination of extinction falling abruptly at the edge of the inner
shell, superposed on the gently rising extinction profile of the outer shell, does mimic
the effect of the trough, so it is feasible that this is one possible explanation. The
other explanation is that dust is genuinely destroyed in a narrow annulus around the
inner shell. Other more exotic causes not involving dust destruction could also
be considered, such as magneto-hydrodynamic interactions between the wind and large-scale 
magnetic fields, as explored by \citet{vanMarle2014} and \citet{Albertazzi2014}.

\subsection{Origin of the dust wave feature}
The trough in extinction appears to separate the two plateaux of dust extinction
which correspond to the inner and outer shells. However, the extinction trough is 
not obvious in the electron density profile (Fig.~\ref{fig:rimsect4}) as this is
steeply declining in this region; nor is the trough apparent in the dust-to-gas map 
(Fig.~\ref{fig:dust_gas}) or the 1D profile (Fig.~\ref{fig:rimsectNh}). Thus one 
interpretation of this feature is that it corresponds to a short decrease in the 
mass loss rate (both dust and gas) during the AGB phase. For a typical AGB expansion
velocity of $\sim$10 kms$^{-1}$, this variation in mass loss would have occurred 
around 3500 yrs ago (assumed distance 1400 pc), for a period of a few hundred years, 
which seems recent relative to the assumed age of the PN outer shell of $\sim$2000 yrs 
\citep{Sabbadin2004}, and older for the halo. Conversely the elevated extinction
just inside the rim of the inner shell (He~II ionization front) could indicate
an earlier increase in the dust mass loss and is also associated with an increase
in the gas density (Fig.~\ref{fig:rimsectNh}). The models of colliding winds of PNe 
show that the shock propagating into the slow AGB wind tends to dilute any strong 
density contrasts (cf., \citet{Steffen2008}) so these features may have been much
stronger and narrower before the PN phase. 

% This needs checking and elaborating.

The presence of two plateaux of dust extinction -- inner shell and outer shell --
separated by a trough is suggestive of a dust destruction mechanism across
the inner-outer shell boundary. The classical mechanism to destroy dust is
sputtering by energetic particles \citep{Barlow1978a}, such as in shocks as 
considered in detail for supernovae by \citet{Bocchio2014}, or by grain-grain 
collisions \citep{Barlow1978b}. However, there is no observational evidence for 
a shock in this region - as for example from broadening or wings to line profiles 
or kinematic discontinuity (\citet{Tsamis2008}; \citet{Sabbadin2004}; 
\citet{Hyung2014}) nor any indication of an increase in ionization through extra 
heating (e.g., enhanced He~II emission or other high ionization lines). Velocities 
in excess of 50 kms$^{-1}$ are typically required to sputter dust, especially since 
NGC~7009 shows evidence for silicate dust which requires 
higher velocities for sputtering than carbonaceous dust \citep{Bocchio2014}.
Based on Fig. 5 of \citet{Sabbadin2004}, the expansion velocity in the vicinity of 
the trough is only $\sim$20 kms$^{-1}$ making significant thermal sputtering by 
protons and heavier atoms unlikely.

The ionization conditions change at the inner/outer shell boundary, in particular the 
He~II emission prominent within the main shell is not present outside implying a 
confinement of the highest ionization gas ($h\nu >$ 54eV) within the main shell.
The X-ray maps of NGC~7009, 
\citet{Guerrero2002} from XMM-Newton and, in particular, from \citet{Kastner2012} for 
Chandra, show diffuse million degree gas confined within the main shell. The spatial 
resolution of the Chandra maps is $\sim$0.5$''$ and the Chandra ACIS soft-band 
image in \citet{Kastner2012} does not show the 10\% contour beyond the main shell.
The X-ray photons could be responsible for some dust processing, but grain heating 
is the major effect of an X-ray flux \citep{Weingartner2006}. The hot, X-ray emitting 
diffuse gas could be responsible for some dust destruction through thermal sputtering 
(\citet{Draine1979} and \citet{Barlow1978b}). However, the crest of higher extinction 
inside the ionization front would be expected to be subject to stronger interaction 
with hot gas, yet remains as a significant peak in extinction suggesting that the 
X-ray emitting gas cannot account for the presence of the trough feature.

If some dust destruction is associated with the trough feature, an estimate of the 
fraction of dust destroyed can be derived by subtracting a smoothed version of the 
extinction map from the map (Fig.~\ref{fig:final_cmap}). The extinction trough 
corresponds to a drop of $c$ of $\sim$0.03. Depending on the smoothing kernel 
applied (Gaussians with FWHM in the range 1.4 to 3.3$''$ were tried), a range
of 3-5\% reduction in the total extinction over the whole nebula is indicated
for this feature. This is very low level and tends to indicate that any
destruction of dust from the previous AGB phase by the PN phase, which
is attributable to the presence of the trough feature, is minimal. 

% The two wind model of PNe formation predicts the presence of a contact discontinuity 
% between the fast, low density wind of the hot central star and slow, high density
% AGB wind and this is where a narrow annulus of X-ray emission occurs \citep{Akashi2006}.
% Allowing for the presence of heat conduction across the contact discontinuity
% broadens this annulus \citep{Steffen2008} and can lead to a more homogeneously filled
% X-ray bubble. 

% \subsection{Reddening curve?}
% The $c$ map determined from the the 10s H$\alpha$ and 120s H$\beta$ images is
% higher by $\sim$0.02 on average over the innner shell than the $c$ map computed from
% H$\alpha$ and the Paschen 9 line. This may be a hint for 
% a reddening law that is steeper with wavelength than the Seaton ISM law with $R$ = 3.1
% (i.e., $R < 3.1$), but the likely flux calibration errors are $\pm$5\% in 
% this wavelength range and it is to be noted that the Paschen 10 and 11 lines are 
% brighter with respect to Paschen 9 by around 10\%.
% ! Computing $c$ from P9/H$\beta$ with R=2.0, gives $c$ lower by $\sim$0.05.
% Possible explanation of extinction torugh could be lower R.

\subsection{Reflection nebula origin of the extinction trough?}
If the elevated extinction observed over the central zone of NGC~7009 is caused by 
a shell of dust at the rim of the volume, as suggested by the geometrical gas and 
dust model in Sect. \ref{sect:Model}, then the lines of sight at the projected
edge of the inner shell would be expected to show higher extinction as a longer dust
column is probed. However, the opposite is the case and the extinction is lower
(Figs. \ref{fig:rimsectc} and \ref{fig:rimsect4}). Assuming that the extinction 
increase measured inside the inner shell relative to the outer
shell (c $\sim$ 0.024, $E_{B-V}$ 0.016 mag.) represents the extinction derived from an
annular dust shell around the the inner shell, the expected tangential increase in
line of sight extinction at its edge can be estimated. For a simple geometrical annulus 
of width 0.5$''$ around the inner shell, the tangent through the prolate shell has an 
extinction increased by up to 9$\times$ (13$\times$ for a oblate form); these figures 
are corroborated by test vectors placed at various positions around the inner shell across 
the trough in the extinction image (Fig.~\ref{fig:final_cmap}). 
% oblate A=B and C<A
%prolate A=B and C>A

The suggestion then arises if the trough in the $c$ map could arise from scattering of 
radiation from the bright inner nebula by a dust shell causing 'blueing' of the 
spectra and hence lower measured extinction. The reduction in extinction from the 
hypothesized dust shell to the measured value in the trough suggests 
a range of $E_{B-V}$ of 0.3 -- 0.4 mag. Such values are 
not untypical of the 'blueing' of reflection nebulae, e.g., $E_{B-V}$ change
$\lesssim$ 0.5 mag. for the reflection nebula NGC~1435, Merope \citep{Hanner1971}. The 
geometrical situation of scattering of radiation from an extended central source 
in NGC~7009 is, however, very distinct from the scattering of radiation from a single star
in NGC~1435 and the masses of gas and dust in a PN are much lower than in typical
reflection nebulae around young stars. A measurement of the linear polarization 
across the trough could confirm the role of dust scattering in this region; 
so far only aperture polarimetry over the outer halo has been 
performed in NGC~7009 \citep{Leroy1986}, with measured values of $\sim$0.5 -- 1\%.
Also mapping of the thermal dust emission could determine the presence of a thick
dust shell in this vicinity.
 
% Try to get a reverse shock velocity and then see if this can destroy Si grains
% Mass loss rate is quite low 2x10$^{-9}$ 

% Based on Sabbaddin Fig. 5, expansion velocity at trough closest to central
% star (offset 5.2'') is 18 km/s. This translates to 1900 years for constant
% velocity (Sabbaddin et al. gives 1650 yars for main shell kinematic age. 
% For a 10 km/s wind this is only 3400 years. 
% In Vassiliadis and Wood Log T = 4.9, Log L = 3.7 would correspond to 
% a 1.5 M sun star at 6000 years.

\section{Conclusions}
Moderately extended planetary nebulae are well suited to integral field spectroscopy with 
an IFU as large as MUSE in its wide field mode, which is capable of covering most of the 
bright emission in a single exposure. The MUSE spaxel scale allied with good seeing 
observations can reveal a wealth of structures in a range of emission lines useful for 
diagnostics of extinction, electron temperature and density, ionization stratification
and chemical abundances over the whole nebula surface. Such data are ideally suited to 
detailed comparison with images and spectra at other wavelengths, such as in the infrared, 
mm/sub-mm and radio.

Analysis of the Balmer (H$\beta$ and H$\alpha$) and brighter Paschen series (P9--P12) 
lines have been described based on MUSE science verification observations of the well 
studied PN NGC~7009. The extinction map reveals a wealth of structures from large
scale shells to individual morphological features. Internal dust extinctions in 
typical PNe have never been mapped before, except for the extreme cases of NGC~7027 
\citep{Walton1988} and NGC~6302 \citep{Matsuura2005}, due to the gas and dust column 
densities being more than an order magnitude smaller than those in typical HII regions.
This result for NGC~7009 was unexpected and attests to the sensitivity of 2D mapping of 
extinction with MUSE.

A map of $A_V/N_{\rm H}$ has been derived from comparison of the extinction map with 
the H line emission for the first time in a PN. An unusual
feature in the extinction map, of a crest and trough at the outer edge of the inner
emission shell, was investigated to determine if it could be connected with dust
reprocessing within the ionized nebula. No convincing evidence was found that it is 
not simply an intrinsic feature of the dust ejected in the antecedent AGB phase.

\begin{acknowledgements}
We would like to thank the MUSE science verification team for the very successful conduct of 
the observations and the whole MUSE team, led by Roland Bacon, for providing such a high 
fidelity IFU ideally matched to spectroscopy of Galactic planetary nebulae. We also 
thank the anonymous referee for comments and suggestions.
\end{acknowledgements}

% WARNING
%-------------------------------------------------------------------
% Please note that we have included the references to the file aa.dem in
% order to compile it, but we ask you to:
%
% - use BibTeX with the regular commands:
\bibliographystyle{aa} % style aa.bst
\bibliography{N7009} % your references Yourfile.bib

\begin{thebibliography}{53}
\expandafter\ifx\csname natexlab\endcsname\relax\def\natexlab#1{#1}\fi

\bibitem[{{Albertazzi} {et~al.}(2014){Albertazzi}, {Ciardi}, {Nakatsutsumi},
  {Vinci}, {B{\'e}ard}, {Bonito}, {Billette}, {Borghesi}, {Burkley}, {Chen},
  {Cowan}, {Herrmannsd{\"o}rfer}, {Higginson}, {Kroll}, {Pikuz}, {Naughton},
  {Romagnani}, {Riconda}, {Revet}, {Riquier}, {Schlenvoigt}, {Skobelev},
  {Faenov}, {Soloviev}, {Huarte-Espinosa}, {Frank}, {Portugall}, {P{\'e}pin},
  \& {Fuchs}}]{Albertazzi2014}
{Albertazzi}, B., {Ciardi}, A., {Nakatsutsumi}, M., {et~al.} 2014, Science,
  346, 325

\bibitem[{{Aller} \& {Czyzak}(1977)}]{Aller1977}
{Aller}, L.~H. \& {Czyzak}, S.~J. 1977, \pasp, 89, 612

\bibitem[{{Aller} \& {Epps}(1975)}]{Aller1975}
{Aller}, L.~H. \& {Epps}, H.~W. 1975, \apj, 197, 175

\bibitem[{{Balick} {et~al.}(1998){Balick}, {Alexander}, {Hajian}, {Terzian},
  {Perinotto}, \& {Patriarchi}}]{Balick1998}
{Balick}, B., {Alexander}, J., {Hajian}, A.~R., {et~al.} 1998, \aj, 116, 360

\bibitem[{{Barker}(1983)}]{Barker1983}
{Barker}, T. 1983, \apj, 267, 630

\bibitem[{{Barlow}(1978{\natexlab{a}})}]{Barlow1978b}
{Barlow}, M.~J. 1978{\natexlab{a}}, \mnras, 183, 397

\bibitem[{{Barlow}(1978{\natexlab{b}})}]{Barlow1978a}
{Barlow}, M.~J. 1978{\natexlab{b}}, \mnras, 183, 367

\bibitem[{{Bocchio} {et~al.}(2014){Bocchio}, {Jones}, \&
  {Slavin}}]{Bocchio2014}
{Bocchio}, M., {Jones}, A.~P., \& {Slavin}, J.~D. 2014, \aap, 570, A32

\bibitem[{{Bohigas} {et~al.}(1994){Bohigas}, {Lopez}, \&
  {Aguilar}}]{Bohigas1994}
{Bohigas}, J., {Lopez}, J.~A., \& {Aguilar}, L. 1994, \aap, 291, 595

\bibitem[{{Czyzak} \& {Aller}(1979)}]{Czyzak1979}
{Czyzak}, S.~J. \& {Aller}, L.~H. 1979, \mnras, 188, 229

\bibitem[{{Draine}(1985)}]{Draine1985}
{Draine}, B.~T. 1985, \apjs, 57, 587

\bibitem[{{Draine}(2011)}]{Draine2011}
{Draine}, B.~T. 2011, {Physics of the Interstellar and Intergalactic Medium}

\bibitem[{{Draine} \& {Lee}(1984)}]{DraineLee1984}
{Draine}, B.~T. \& {Lee}, H.~M. 1984, \apj, 285, 89

\bibitem[{{Draine} \& {Salpeter}(1979)}]{Draine1979}
{Draine}, B.~T. \& {Salpeter}, E.~E. 1979, \apj, 231, 438

\bibitem[{{Fang} \& {Liu}(2011)}]{Fang2011}
{Fang}, X. \& {Liu}, X.-W. 2011, \mnras, 415, 181

\bibitem[{{Fang} \& {Liu}(2013)}]{Fang2013}
{Fang}, X. \& {Liu}, X.-W. 2013, \mnras, 429, 2791

\bibitem[{{Frew}(2008)}]{Frew}
{Frew}, D.~J. 2008, PhD thesis, Department of Physics, Macquarie University,
  NSW 2109, Australia

\bibitem[{{Fried}(1966)}]{Fried1966}
{Fried}, D.~L. 1966, Journal of the Optical Society of America (1917-1983), 56,
  1372

\bibitem[{{Gon{\c c}alves} {et~al.}(2003){Gon{\c c}alves}, {Corradi},
  {Mampaso}, \& {Perinotto}}]{Goncalves2003}
{Gon{\c c}alves}, D.~R., {Corradi}, R.~L.~M., {Mampaso}, A., \& {Perinotto}, M.
  2003, \apj, 597, 975

\bibitem[{{Green} {et~al.}(2015){Green}, {Schlafly}, {Finkbeiner}, {Rix},
  {Martin}, {Burgett}, {Draper}, {Flewelling}, {Hodapp}, {Kaiser}, {Kudritzki},
  {Magnier}, {Metcalfe}, {Price}, {Tonry}, \& {Wainscoat}}]{Green2015}
{Green}, G.~M., {Schlafly}, E.~F., {Finkbeiner}, D.~P., {et~al.} 2015, \apj,
  810, 25

\bibitem[{{Gudennavar} {et~al.}(2012){Gudennavar}, {Bubbly}, {Preethi}, \&
  {Murthy}}]{Gudennavar2012}
{Gudennavar}, S.~B., {Bubbly}, S.~G., {Preethi}, K., \& {Murthy}, J. 2012,
  \apjs, 199, 8

\bibitem[{{Guerrero} {et~al.}(2002){Guerrero}, {Gruendl}, \&
  {Chu}}]{Guerrero2002}
{Guerrero}, M.~A., {Gruendl}, R.~A., \& {Chu}, Y.-H. 2002, \aap, 387, L1

\bibitem[{{Hanner}(1971)}]{Hanner1971}
{Hanner}, M.~S. 1971, \apj, 164, 425

\bibitem[{{Howarth}(1983)}]{Howarth1983}
{Howarth}, I.~D. 1983, \mnras, 203, 301

\bibitem[{{Hyung} \& {Aller}(1995{\natexlab{a}})}]{Hyung1995A}
{Hyung}, S. \& {Aller}, L.~H. 1995{\natexlab{a}}, \mnras, 273, 973

\bibitem[{{Hyung} \& {Aller}(1995{\natexlab{b}})}]{Hyung1995B}
{Hyung}, S. \& {Aller}, L.~H.~H. 1995{\natexlab{b}}, \mnras, 273, 958

\bibitem[{{Hyung} {et~al.}(2014){Hyung}, {Lee}, \& {Sung}}]{Hyung2014}
{Hyung}, S., {Lee}, S.-J., \& {Sung}, E.-C. 2014, \apss, 352, 71

\bibitem[{{Kastner} {et~al.}(2012){Kastner}, {Montez}, {Balick}, {Frew},
  {Miszalski}, {Sahai}, {Blackman}, {Chu}, {De Marco}, {Frank}, {Guerrero},
  {Lopez}, {Rapson}, {Zijlstra}, {Behar}, {Bujarrabal}, {Corradi}, {Nordhaus},
  {Parker}, {Sandin}, {Sch{\"o}nberner}, {Soker}, {Sokoloski}, {Steffen},
  {Ueta}, \& {Villaver}}]{Kastner2012}
{Kastner}, J.~H., {Montez}, Jr., R., {Balick}, B., {et~al.} 2012, \aj, 144, 58

\bibitem[{{Leroy} {et~al.}(1986){Leroy}, {Le Borgne}, \& {Arnaud}}]{Leroy1986}
{Leroy}, J.~L., {Le Borgne}, J.~F., \& {Arnaud}, J. 1986, \aap, 160, 171

\bibitem[{{Liu} \& {Danziger}(1993)}]{Liu1993}
{Liu}, X.-W. \& {Danziger}, J. 1993, \mnras, 263, 256

\bibitem[{{Liu} {et~al.}(1995){Liu}, {Storey}, {Barlow}, \& {Clegg}}]{Liu1995}
{Liu}, X.-W., {Storey}, P.~J., {Barlow}, M.~J., \& {Clegg}, R.~E.~S. 1995,
  \mnras, 272, 369

\bibitem[{{Mathis} {et~al.}(1977){Mathis}, {Rumpl}, \&
  {Nordsieck}}]{Mathis1977}
{Mathis}, J.~S., {Rumpl}, W., \& {Nordsieck}, K.~H. 1977, \apj, 217, 425

\bibitem[{{Matsuura} {et~al.}(2005){Matsuura}, {Zijlstra}, {Molster}, {Waters},
  {Nomura}, {Sahai}, \& {Hoare}}]{Matsuura2005}
{Matsuura}, M., {Zijlstra}, A.~A., {Molster}, F.~J., {et~al.} 2005, \mnras,
  359, 383

\bibitem[{{Mendez} {et~al.}(1992){Mendez}, {Kudritzki}, \&
  {Herrero}}]{Mendez1992}
{Mendez}, R.~H., {Kudritzki}, R.~P., \& {Herrero}, A. 1992, \aap, 260, 329

\bibitem[{{Mendez} {et~al.}(1988){Mendez}, {Kudritzki}, {Herrero}, {Husfeld},
  \& {Groth}}]{Mendez1988}
{Mendez}, R.~H., {Kudritzki}, R.~P., {Herrero}, A., {Husfeld}, D., \& {Groth},
  H.~G. 1988, \aap, 190, 113

\bibitem[{{Perinotto} {et~al.}(1998){Perinotto}, {Kifonidis}, {Schoenberner},
  \& {Marten}}]{Perinotto1998}
{Perinotto}, M., {Kifonidis}, K., {Schoenberner}, D., \& {Marten}, H. 1998,
  \aap, 332, 1044

\bibitem[{{Phillips} {et~al.}(2010){Phillips}, {Cuesta}, \&
  {Ramos-Larios}}]{Phillips2010}
{Phillips}, J.~P., {Cuesta}, L.~C., \& {Ramos-Larios}, G. 2010, \mnras, 409,
  881

\bibitem[{{Planck Collaboration} {et~al.}(2015){Planck Collaboration},
  {Arnaud}, {Atrio-Barandela}, {Aumont}, {Baccigalupi}, {Banday}, {Barreiro},
  {Battaner}, {Benabed}, {Benoit-L{\'e}vy}, {Bernard}, {Bersanelli},
  {Bielewicz}, {Bonaldi}, {Bond}, {Borrill}, {Bouchet}, {Buemi}, {Burigana},
  {Cardoso}, {Casassus}, {Catalano}, {Cerrigone}, {Chamballu}, {Chiang},
  {Colombi}, {Colombo}, {Couchot}, {Crill}, {Curto}, {Cuttaia}, {Davies},
  {Davis}, {de Bernardis}, {de Rosa}, {de Zotti}, {Delabrouille}, {Dickinson},
  {Diego}, {Donzelli}, {Dor{\'e}}, {Dupac}, {En{\ss}lin}, {Eriksen}, {Finelli},
  {Frailis}, {Franceschi}, {Galeotta}, {Ganga}, {Giard}, {Gonz{\'a}lez-Nuevo},
  {G{\'o}rski}, {Gregorio}, {Gruppuso}, {Hansen}, {Harrison}, {Hildebrandt},
  {Hivon}, {Holmes}, {Hora}, {Hornstrup}, {Hovest}, {Huffenberger}, {Jaffe},
  {Jones}, {Juvela}, {Keih{\"a}nen}, {Keskitalo}, {Kisner}, {Knoche}, {Kunz},
  {Kurki-Suonio}, {L{\"a}hteenm{\"a}ki}, {Lamarre}, {Lasenby}, {Lawrence},
  {Leonardi}, {Leto}, {Lilje}, {Linden-V{\o}rnle}, {L{\'o}pez-Caniego},
  {Mac{\'{\i}}as-P{\'e}rez}, {Maffei}, {Maino}, {Mandolesi}, {Martin}, {Masi},
  {Massardi}, {Matarrese}, {Mazzotta}, {Mendes}, {Mennella}, {Migliaccio},
  {Miville-Desch{\^e}nes}, {Moneti}, {Montier}, {Morgante}, {Mortlock},
  {Munshi}, {Murphy}, {Naselsky}, {Nati}, {Natoli}, {Noviello}, {Novikov},
  {Novikov}, {Pagano}, {Pajot}, {Paladini}, {Paoletti}, {Peel}, {Perdereau},
  {Perrotta}, {Piacentini}, {Piat}, {Pietrobon}, {Plaszczynski},
  {Pointecouteau}, {Polenta}, {Popa}, {Pratt}, {Procopio}, {Prunet}, {Puget},
  {Rachen}, {Reinecke}, {Remazeilles}, {Ricciardi}, {Riller}, {Ristorcelli},
  {Rocha}, {Rosset}, {Roudier}, {Rubi{\~n}o-Mart{\'{\i}}n}, {Rusholme},
  {Sandri}, {Savini}, {Scott}, {Spencer}, {Stolyarov}, {Sutton}, {Suur-Uski},
  {Sygnet}, {Tauber}, {Terenzi}, {Toffolatti}, {Tomasi}, {Trigilio},
  {Tristram}, {Trombetti}, {Tucci}, {Umana}, {Valiviita}, {Van Tent}, {Vielva},
  {Villa}, {Wade}, {Wandelt}, {Zacchei}, {Zijlstra}, \& {Zonca}}]{Planck2015}
{Planck Collaboration}, {Arnaud}, M., {Atrio-Barandela}, F., {et~al.} 2015,
  \aap, 573, A6

\bibitem[{{Rodr{\'{\i}}guez} \& {G{\'o}mez}(2007)}]{Rodriguez2007}
{Rodr{\'{\i}}guez}, L.~F. \& {G{\'o}mez}, Y. 2007, \rmxaa, 43, 173

\bibitem[{{Rubin} {et~al.}(2002){Rubin}, {Bhatt}, {Dufour}, {Buckalew},
  {Barlow}, {Liu}, {Storey}, {Balick}, {Ferland}, {Harrington}, \&
  {Martin}}]{Rubin2002}
{Rubin}, R.~H., {Bhatt}, N.~J., {Dufour}, R.~J., {et~al.} 2002, \mnras, 334,
  777

\bibitem[{{Sabbadin} {et~al.}(2004){Sabbadin}, {Turatto}, {Cappellaro},
  {Benetti}, \& {Ragazzoni}}]{Sabbadin2004}
{Sabbadin}, F., {Turatto}, M., {Cappellaro}, E., {Benetti}, S., \& {Ragazzoni},
  R. 2004, \aap, 416, 955

\bibitem[{{Schlafly} \& {Finkbeiner}(2011)}]{Schlafly2011}
{Schlafly}, E.~F. \& {Finkbeiner}, D.~P. 2011, \apj, 737, 103

\bibitem[{{Schlegel} {et~al.}(1998){Schlegel}, {Finkbeiner}, \&
  {Davis}}]{Schlegel1998}
{Schlegel}, D.~J., {Finkbeiner}, D.~P., \& {Davis}, M. 1998, \apj, 500, 525

\bibitem[{{Seaton}(1979)}]{Seaton1979}
{Seaton}, M.~J. 1979, \mnras, 187, 73P

\bibitem[{{Steffen} {et~al.}(2008){Steffen}, {Sch{\"o}nberner}, \&
  {Warmuth}}]{Steffen2008}
{Steffen}, M., {Sch{\"o}nberner}, D., \& {Warmuth}, A. 2008, \aap, 489, 173

\bibitem[{{Steffen} {et~al.}(2009){Steffen}, {Esp{\'{\i}}ndola},
  {Mart{\'{\i}}nez}, \& {Koning}}]{Steffen2009}
{Steffen}, W., {Esp{\'{\i}}ndola}, M., {Mart{\'{\i}}nez}, S., \& {Koning}, N.
  2009, \rmxaa, 45, 143

\bibitem[{{Storey} \& {Hummer}(1995)}]{Storey1995}
{Storey}, P.~J. \& {Hummer}, D.~G. 1995, \mnras, 272, 41

\bibitem[{{Tsamis} {et~al.}(2008){Tsamis}, {Walsh}, {P{\'e}quignot}, {Barlow},
  {Danziger}, \& {Liu}}]{Tsamis2008}
{Tsamis}, Y.~G., {Walsh}, J.~R., {P{\'e}quignot}, D., {et~al.} 2008, \mnras,
  386, 22

\bibitem[{{van Marle} {et~al.}(2014){van Marle}, {Cox}, \&
  {Decin}}]{vanMarle2014}
{van Marle}, A.~J., {Cox}, N.~L.~J., \& {Decin}, L. 2014, \aap, 570, A131

\bibitem[{{Walton} {et~al.}(1988){Walton}, {Pottasch}, {Reay}, \&
  {Taylor}}]{Walton1988}
{Walton}, N.~A., {Pottasch}, S.~R., {Reay}, N.~K., \& {Taylor}, A.~R. 1988,
  \aap, 200, L21

\bibitem[{{Weilbacher} {et~al.}(2014){Weilbacher}, {Streicher}, {Urrutia},
  {P{\'e}contal-Rousset}, {Jarno}, \& {Bacon}}]{Weilbacher2014}
{Weilbacher}, P.~M., {Streicher}, O., {Urrutia}, T., {et~al.} 2014, in
  Astronomical Society of the Pacific Conference Series, Vol. 485, Astronomical
  Data Analysis Software and Systems XXIII, ed. N.~{Manset} \& P.~{Forshay},
  451

\bibitem[{{Weingartner} {et~al.}(2006){Weingartner}, {Draine}, \&
  {Barr}}]{Weingartner2006}
{Weingartner}, J.~C., {Draine}, B.~T., \& {Barr}, D.~K. 2006, \apj, 645, 1188

\bibitem[{{Zhang} {et~al.}(2004){Zhang}, {Liu}, {Wesson}, {Storey}, {Liu}, \&
  {Danziger}}]{Zhang2004}
{Zhang}, Y., {Liu}, X.-W., {Wesson}, R., {et~al.} 2004, \mnras, 351, 935

\end{thebibliography}
%
% - join the .bib files when you upload your source files
%-------------------------------------------------------------------

% \begin{thebibliography}{}
% \end{thebibliography}

\end{document}